\documentclass[a4paper,11pt]{scrartcl}

\usepackage{lineno}
\usepackage[utf8]{inputenc}
\usepackage[T1]{fontenc}

\usepackage{amsmath,amsthm,amsopn,xspace,bm}

\usepackage[english]{babel}

\usepackage{graphicx}

\usepackage[babel]{microtype}

\usepackage{bera}
\usepackage[charter]{mathdesign}
\usepackage[colorinlistoftodos,shadow]{todonotes}

\begin{document}


\subject{Short Note}

\title{On the Proper Setup of the Double Mach Reflection as a Test Case for
  the Resolution of Gas Dynamics Codes}

\author{Friedemann Kemm\footnote{Brandenburg University of Technology,
    Platz der Deutschein Einheit 1, 03046 Cottbus, Germany,
    kemm@math.tu-cottbus.de}}

\publishers{\rule{0pt}{1pt}\\\textbf{Keywords:} Double Mach
  reflection, high speed flow, high resolution scheme, numerical test
  case}

\maketitle

\paragraph*{MSC 2010:} 76J20, 
  76L05,  
  76M99  

\begin{abstract}
  This note discusses the initial and boundary conditions as well as
  the size of the computational domain for the double Mach reflection
  problem when set up as a test for the resolution of an Euler scheme
  for gas dynamics. 
\end{abstract}

\section{The double Mach reflection}
\label{sec:double-mach-refl}

\begin{figure}
  \centering
  \includegraphics[width=.8\linewidth]{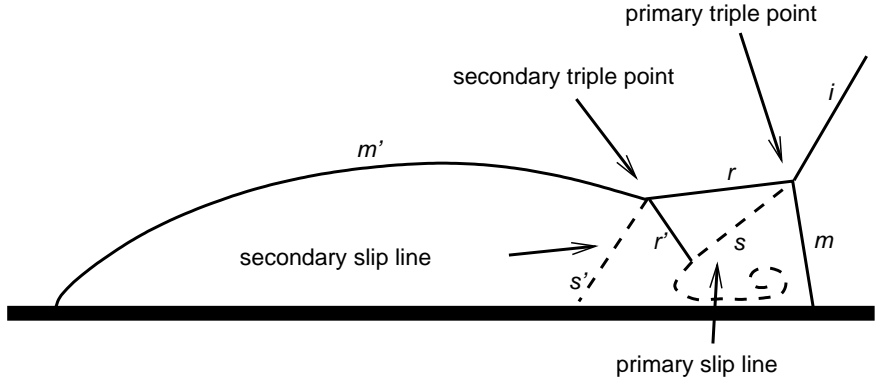}
  \caption{Sketch of the double Mach reflection problem. The bottom
    line represents the ramp.} 
  \label{fig:dmr1sketch}
\end{figure}

A famous test for the quality of a Riemann solver is the Double Mach
reflection problem. It was suggested by Woodward and Colella~\cite{WC}
as a benchmark for Euler codes.
An analytical treatment is found in~\cite{double-mach}
and~\cite{ben-dor-buch} and the references therein, while experimental
results are presented in~\cite{gvozdeva_double_1968} and also
in~\cite[pp.~152~and~168]{ben-dor-buch}. The problem consists of a
shock front that hits a ramp which is inclined by~30~degrees. When the
shock runs up the ramp, a self similar shock structure with two triple
points evolves. The situation is sketched out in
Figure~\ref{fig:dmr1sketch}. 
To simplify the graphical representation, the coordinate system is
aligned with the ramp\,--\,as done for the numerical tests. In the
primary triple point, the incident shock~\(i\), the mach stem~\(m\),
and the reflected shock~\(r\) meet. In the double mach configuration,
the reflected shock breaks up forming a secondary triple point with
the reflected shock~\(r\), a secondary (bowed) mach stem~\(m'\), and a
secondary reflected shock~\(r'\). From both triple points, slip lines
emanate. The reflected shock~\(r'\) hits the primary slip line~\(s\)
causing a curled flow structure, the resolution of which may serve as
an indicator for the resolution of a numerical scheme. But---as was
already stated by Woodward and Colella---the main challenge for a high
resolution scheme is to resolve the secondary slip line~\(s'\). Being
a rather weak feature, it is hardly visible in a density plot or a
plot of any velocity component. According to Woodward and
Colella~\cite{WC}, the secondary slip~\(s'\) line can be best observed
in the vertical momentum. Thus, throughout this paper we present the
results for the vertical momentum~\(\rho v\).

\section{The problem: two issues}
\label{sec:problem:-two-issues}

\begin{figure}
  \centering
  \includegraphics[width=.8\linewidth]{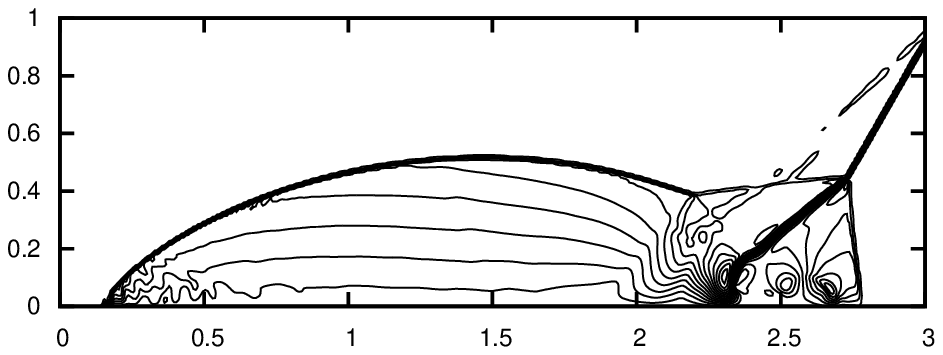}\\
  \includegraphics[width=.8\linewidth]{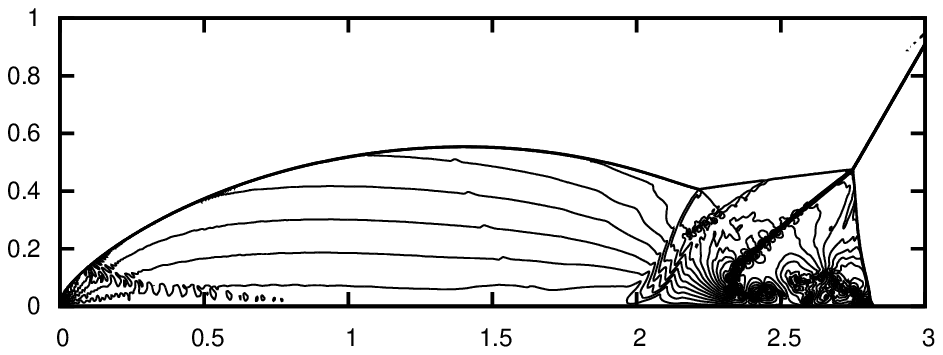}
  \caption{Numerical artifact near the secondary slip line in the
    standard setting with~\(\Delta x = \Delta y = 1/120\) (top)
    and~\(\Delta x = \Delta y = 1/480\) (bottom).}
  \label{fig:disturbance}
\end{figure}

For our numerical tests, we use the setting as described
in~\cite{WC}. We start with the shock already on the ramp and rotate
the coordinate system and, so that the computational grid is aligned
with the ramp.  The initial conditions left and right of the shock are
\begin{align*}
  \rho_l & = 8.0\;, & (\rho u)_l & = 57.1597\;, & (\rho v)_l & =
  -33.0012\;, & E_l & = 563.544\;, \\
  \rho_r & = 1.4\;, & (\rho u)_r & = 0\;, & (\rho v)_r & = 0\;, & E_r
  & = 2.5\;.
\end{align*}
The initial shock hits the bottom of the computational domain at~\(x_0
= 1/6\). Usually the computational domain is chosen
as~\([0,4]\times[0,1]\) and the results are presented
for~\(t=0.2\). At the bottom, we employ solid wall conditions, at the
right boundary outflow. At all other boundaries we use
Dirichlet-conditions, which are set to the physical values. 

Unfortunately, as the results in Figure~\ref{fig:disturbance} show,
there are severe disturbances of the flow close to the secondary slip
line. Depending on the scheme and the grid resolution, it is difficult
to distinguish between slip line and numerical artifact.

As Figure~\ref{fig:disturbance} shows, this setting results in some
numerical artifacts disturbing the secondary reflected shock~\(r'\)
and the region between the secondary reflected shock and the secondary
slip line. At lower resolutions (top picture), the artifacts are not
distinguishable from the secondary slip line~\(s'\). While Woodward
and Colella~\cite{WC} blame this effect to the under-resolved shock in
the initial condition, Rider et~al.~\cite{pre05191152} argue that the
under-resolved shock in the boundary condition for the upper boundary
is responsible for it. Both are partially right and partially wrong:
\begin{figure}
  \centering
  \includegraphics[width=.8\linewidth]{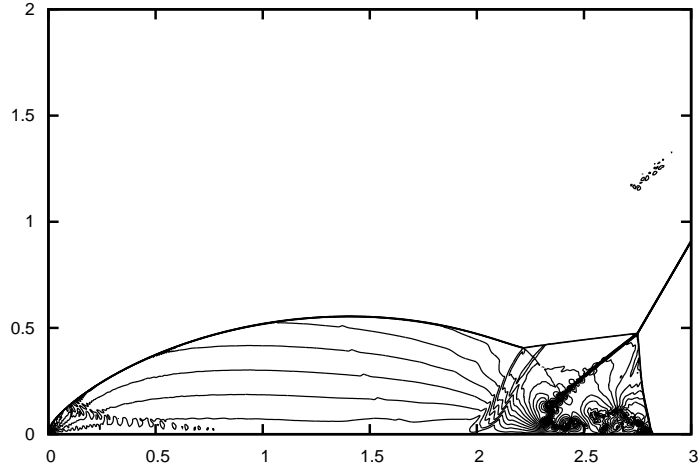}
  \caption{Double Mach reflection computed on~\([0,4]\times[0,2]\).}
  \label{fig:y2}
\end{figure}
In Figure~\ref{fig:y2}, we show results for the double Mach reflection
computed on~\([0,4]\times[0,2]\) instead of~\([0,4]\times[0,1]\). It
can be seen that what in Figure~\ref{fig:disturbance} seemed to be one
(kinked) phenomenon in fact are two artifacts: one arising from the
shock position at the upper boundary---it shows up as a slight
disturbance above the shock close to the right boundary and as a
slight disturbance a little bit left of the secondary slip line---and
one that follows the shock at a certain distance (and slightly to the
right of the secondary slip line), indicating that it results from the
initial condition. This means that there indeed is an artifact arising
from the initial condition (hypothesis by Woodward and Colella) and an
artifact arising from the boundary condition (hypothesis by Rider
et~al.).

In the following, we will investigate both hypotheses by means of
numerical tests with different settings. This will give us some hints
on the proper use of the double Mach reflection as a test case for
Euler codes.

\section{The numerical environment for the tests}
\label{sec:numer-envir-tests}

To set up tests in order to investigate the hypotheses by Rider
et~al.\ and by Woodward and Colella, one has to make sure that the
grid resolution is the only variable parameter in the numerical
scheme. Besides this, the size of the computational domain and the
initial and the boundary conditions may vary. But the basic features
of the method have to be fixed, including Riemann solver, grid
structure, basic approach (finite differences, finite volumes,
discontinuous Galerkin,\dots), reconstruction techniques, limiters,
time scheme etc. In this study, we resort to finite volumes on a
uniform equidistant Cartesian grid with~\(\Delta x = \Delta y\). The
basic scheme uses wave propagation according to LeVeque~\cite{leveque}
with algebraic limiting and Roe with Harten-Hyman entropy
fix~\cite{harten-hyman}. Thus, it is a second order TVD-scheme. The
second order corrections are applied also for the corner fluxes. As
limiter, we employ the mixed use of CFL-Superbee and Superpower as
described in~\cite{limiter}, modified for nonlinear waves according to
Jeng and Payne~\cite{jeng-tvd} as described
in~\cite{kemm-santiago}. The code used for the examples is
clawpack~\cite{clawpack}. As for Figure~\ref{fig:disturbance}, we do
not show the entire computational domain. We restrict the
\(x\)-direction to~\([0,3]\) or, in Figure~\ref{fig:ext}, show a
close-up of the region of interest: the region containing the triple
points. As already mentioned, the quantity shown is always the
vertical momentum~\(\rho v\). And, if not stated otherwise, the grid
resolution is~\(\Delta x = \Delta y = 1/480\).

\section{Hypothesis by Rider et al.}
\label{sec:hypothesis-rider-et}

According to Rider et~al.~\cite{pre05191152}, the under-resolved shock
in the boundary condition forces a reflection phenomenon which can be
avoided if the shock is smeared over three grid cells. The tests in
Figure~\ref{fig:disturbance} were executed with such a
setting. Obviously, for high grid resolutions, this is not
sufficient. Even smearing over seven grid cells would not give the
desired results. Furthermore, it is not at all clear if a linear
smearing would be appropriate or if a more complex strategy has to be
chosen.

Another idea would be to switch the boundary conditions at the upper
boundary from Dirichlet to oblique first-order extrapolation. 
\begin{figure}
  \centering
\includegraphics[width=.5\linewidth]{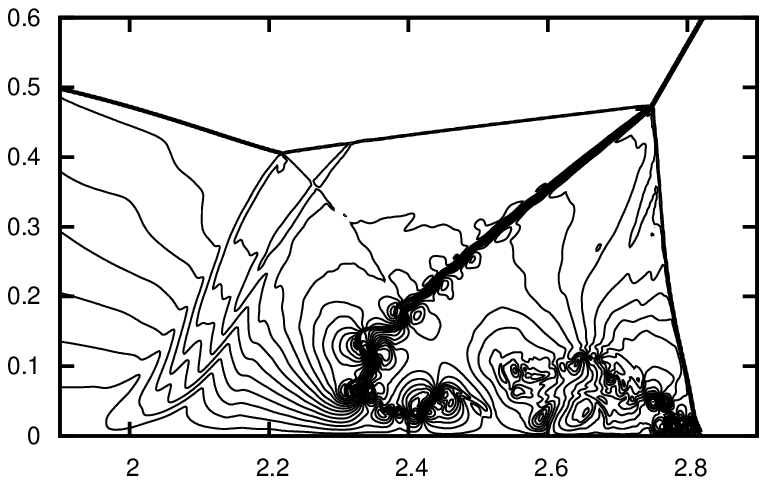}\hspace{-2em}\includegraphics[width=.5\linewidth]{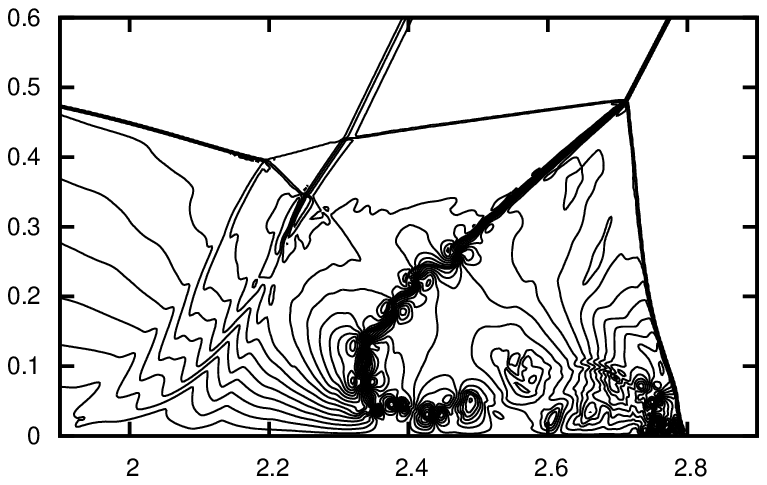}
\caption{Effect of boundary condition at top boundary: Dirichlet
  (left) and oblique extrapolation (right).}
  \label{fig:ext}
\end{figure}
%
Since the angle of the shock to the vertical grid lines is 30°,
for~\(\Delta x = \Delta y\) this
can be easily achieved by setting
\begin{equation}
  \label{eq:1}
  \bm q_{i,m+1} = \bm q_{i-1,m-1}\;, \qquad \bm q_{i,m+2} = \bm
  q_{i-1,m}\;, 
\end{equation}
where~\(m\) is the number of grid cells in the~\(y\)-direction. As
Figure~\ref{fig:ext} indicates, the reflection is no longer
seen. However, the shock runs, also the angle of the primary shock
changes over time and, thus, changes the structure of the solution. It
is no longer self-similar. Furthermore, the extrapolation enhances the
resolution of the other numerical artifact and, thus, deteriorates the
quality of the solution even more.

In summary, the best strategy to address the reflection from the top
boundary is---as shown for Figure~\ref{fig:y2} in the previous
section---to enlarge the computational domain in
the~\(y\)-direction. This way, one can make sure that it preserves the
self-similar structure of the solution and, at the same time, does not
interact with the secondary slip line. It hits the secondary Mach
stem~\(m'\) far left of the interesting part of the solution.

\section{Hypothesis by Woodward and Colella}
\label{sec:hypoth-woodw-colella}

While the artifact arising from the boundary condition can easily be
removed from the region of interest, this is not true for the artifact
arising from the initial condition which is part of the self-similar
numerical---not of the physical---structure of the solution. According
to Woodward and Colella~\cite{WC}, the artifact arising from the
initial condition results from the fact that a shock which does not
start at a grid line, but is somehow projected onto the grid, is
automatically under-resolved. 

\begin{figure}
  \centering
\includegraphics[width=.5\linewidth]{high-supsupmix-y2-sw}\hspace{-2em}\includegraphics[width=.5\linewidth]{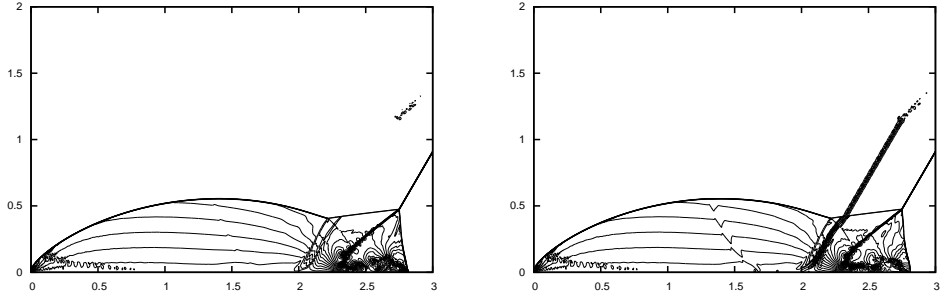}
\caption{Effect of initial condition: with sharp initial shock (left),
  with smeared initial shock (right).}
  \label{fig:smear}
\end{figure}

Therefore, smearing the shock in the initial condition should show
some considerable improvement. However, Figure~\ref{fig:smear} reveals
quite the opposite.
%
%
Essentially, there is no general way to represent a non-grid-aligned
shock without forcing unphysical waves arising from the initial
oblique Riemann problem. It seems even doubtful whether this might
work for any shock capturing scheme, no matter how much information on
the scheme is invested in the construction of the discrete initial
state.

\section{Consequences for the setup of the double Mach reflection as a test}
\label{sec:cons-setup-double}

Although the presented results are gained on a fixed uniform grid, we
can draw some conclusions for both cases: (1) Test with fixed grid
resolution, which is intended to test the quality of the basic scheme;
(2) Test with an adaptive grid, which mainly tests the quality of the
refinement/coarsening strategy.

(1) With fixed grid resolution: The artifact due to the under-resolved
shock in the boundary condition for the upper boundary can be easily
removed from the region if interest by enlarging the computational
domain in the~\(y\)-direction. For~\([0,4]\times[0,2]\) instead
of~\([0,4]\times[0,1]\), the artifact is far left of the secondary
triple point and the secondary slip line.
We have found that it is essentially impossible to get rid of the
artifact due to the under-resolved initial shock.
In order to get information on the quality of the basic numerical
scheme, we need to ensure that the grid is fine enough to separate the
numerical artifact from the secondary slip line, cf.\
Figure~\ref{fig:disturbance}.

(2) With adaptive grid resolution: To test the quality of a grid
refinement/coarsening strategy, it is reasonable to keep the original
setting of Woodward and Colella. A good refinement/coarsening strategy
has to make sure that physical features like the secondary slip line
are refined while numerical artifacts are not. They have to be
invisible in the numerical results, while the secondary slip line has
to be well resolved.

\bibliographystyle{plain} 
\bibliography{limiter,carbuncle,sonst,involutionen}

\end{document}